\documentclass[aip,jcp]{revtex4-1}

\usepackage{bm}
\usepackage{amsmath,amssymb}

\providecommand{\eqnref}[1]{Eq.\ \eqref{#1}}
\renewcommand{\vec}[1]{\ensuremath{\mathbf{#1}}}
\providecommand{\abs}[1]{\lvert#1\rvert}
\providecommand{\transp}{\text{T}}
\providecommand{\ket}[1]{\ensuremath{\lvert#1\rangle}}
\providecommand{\bra}[1]{\ensuremath{\langle #1\rvert}}
\providecommand{\braket}[3]{\ensuremath{
 \langle #1\lvert #2\rvert #3\rangle}}
\providecommand{\mat}[1]{\ensuremath{\bm{\mathsf{#1}}}}
\providecommand{\tj}[6]{\begin{pmatrix} #1 & #2 & #3\\ #4 & #5 & #6
 \end{pmatrix}}
\providecommand{\ito}[2]{\ensuremath{S^{(#1)}_{#2}}}
\providecommand{\itom}[2]{\ensuremath{\mat{S}^{(#1)}_{#2}}}
\providecommand{\bohr}{\ensuremath{\mu_\text{B}}}
\providecommand{\nbohr}{\ensuremath{\mu_\text{N}}}
\DeclareMathOperator{\tr}{Tr}

\begin{document}
\title{NMR chemical shift as analytical derivative of the Helmholtz free energy}
\author{Willem Van den Heuvel}
\author{Alessandro Soncini}
\email{asoncini@unimelb.edu.au}
\affiliation{School of Chemistry, The University of Melbourne, VIC 3010, Australia}

\date{\today}
\begin{abstract}

We present a theory for the temperature-dependent nuclear magnetic shielding
tensor of molecules with arbitrary electronic structure. The theory is a
generalization of Ramsey's theory for closed-shell molecules. The shielding
tensor is defined as a second derivative of the Helmholtz free energy of the
electron system in equilibrium with the applied magnetic field and the nuclear
magnetic moments. This derivative is analytically evaluated and expressed as a
sum over states formula. Special consideration is given to a system with an
isolated degenerate ground state for which the size of the degeneracy and the
composition of the wave functions are arbitrary. In this case the paramagnetic
part of the shielding tensor is expressed in terms of the $g$ and $A$ tensors
of the EPR spin Hamiltonian of the degenerate state. As an illustration of the
proposed theory, we provide an explicit formula for the paramagnetic shift of 
the central lanthanide ion in endofullerenes Ln@C$_{60}$, 
with Ln=Ce$^{3+}$, Nd$^{3+}$, Sm$^{3+}$, Dy$^{3+}$, Er$^{3+}$ and Yb$^{3+}$,
where the ground state can be a strongly spin-orbit coupled icosahedral sextet 
for which the paramagnetic shift cannot be described by previous theories.
\end{abstract}

\maketitle
\section{Introduction}

Nuclear magnetic shielding measured in NMR spectroscopy is caused by the
electrons' response to an externally applied magnetic field. In closed-shell
molecules this response consists of an induced magnetic field caused by
electronic orbital currents, the electron spin playing no role other than
forcing appropriate orbital permutation symmetries in the many-body
wave function.  This, at least, is true when spin-orbit coupling is ignored.
Ramsey, in Ref.\ \onlinecite{Ramsey1950}, presented the quantum mechanical
theory of the shielding tensor for molecules in this regime. 

In open-shell molecules however, there is a second contribution to the
shielding, arising from the permanent magnetic moment associated with the spin
of the unpaired electrons. This contribution is known as the `paramagnetic
shift'.\cite{McConnell1958,Kurland1970,Rinkevicius2003,Moon2004,Soncini2007,
Pennanen2008,Gunne2009,Autschbach2011,VandenHeuvel2012} More generally, a
paramagnetic shift arises from a degenerate electronic state.  When the
degeneracy is weakly split in a magnetic field, the magnetic polarization of
each thermally populated state can be described in terms of {\em permanent}
non-compensating spin (and orbital) currents,\cite{Soncini2007} inducing a net
response field at the nucleus, each state contributing in proportion to its
thermal population.  A general equation for the isotropic paramagnetic shift
was first derived by Kurland and McGarvey.\cite{Kurland1970} Moon and
Patchkovskii derived a formula for the paramagnetic shielding tensor in an
arbitrary Kramers doublet and expressed this formula in terms of the $g$ and
$A$ tensors of the EPR spin Hamiltonian.\cite{Moon2004} Later Pennanen and
Vaara extended the Moon and Patchkovskii theory of paramagnetic shifts to cases
which deviate from a pure spin degeneracy in the lowest order of perturbation
theory in the spin-orbit coupling.\cite{Pennanen2008} 

Although the work of Pennanen and Vaara represents the first comprehensive
effort towards a completely general theory of NMR chemical shielding for
electronically degenerate states in the weak spin-orbit coupling regime, their
approach\cite{Pennanen2008} exposes a few issues: (i) It is developed using
perturbation theory on the thermodynamic internal energy $U$, which appears to
require different perturbation expansions of a same energy level, according to
whether the level is either a weighted addend, or an exponent of the thermal
weight, in the Boltzmann sum-over-states.  (ii) It considers Zeeman and
hyperfine spin Hamiltonians linear in the spin operators, which is strictly
correct only if the degeneracy is no larger than threefold.  (iii) It does not
provide a formulation of the problem for an arbitrary electronic degeneracy,
such as can be found for example in strongly spin-orbit coupled systems. (iv)
Finally, the inclusion of low-lying excited states resulting from weak
splitting of a degeneracy, as described by a zero field splitting Hamiltonian
for pure spin states, is not described in the most general way.

We have recently published a communication in which especially points (iii) and
(iv) were thoroughly discussed and generalized.\cite{VandenHeuvel2012}  In the
present paper we wish to present a theory which offers a rigorous solution to point (i). 
We do this by taking no more than Ramsey's original assumptions \cite{Ramsey1950} and 
applying them to a system where the electrons are in thermal equilibrium.
Thus we are led to identifying the shielding tensor with a second
derivative of the Helmholtz free energy, as in \eqnref{sigma_F}. Then, without
needing further assumptions, we use perturbation theory on the free energy to
derive the expression for the shielding tensor, \eqnref{sigma}. In the second
part of the paper, we show that this expression lends itself easily to a
reformulation of the paramagnetic shielding in a degenerate state in terms of
the spin Hamiltonian $g$ and $A$ parameters (\eqnref{sigmageneral}), which
provides a general solution to point (ii).

\section{Non-degenerate ground state: Shieldings as energy derivatives}

In the following we shall be concerned exclusively with nuclear shielding in
the `solid state limit',\cite{McConnell1958} by which is meant that the
molecular nuclei are fixed in space with respect to the static external
magnetic field. 

In this section we want to recall the essential points of Ramsey's 
theory for a molecule in a non-degenerate ground state.\cite{Ramsey1950}
He observed that the nuclear magnetic moment $\bm{\mu}$ can be treated as a
classical vector because it is much slower in its dynamics than the electrons.
The idea is that as the nuclear moments go about their rotating motion the
electron cloud follows adiabatically; at every moment the electrons are in the
ground state that corresponds to the instantaneous orientation of the nuclear
moments. Hence the components of $\bm{\mu}$ become external parameters in the
electronic Born--Oppenheimer Hamiltonian, as are the nuclear positions and the
applied field $\vec{B}$. The ground state energy will naturally be a function
of $\bm{\mu}$ and $\vec{B}$, and can be expanded in a Taylor series as
\begin{equation}\label{energyexp}
E(\bm{\mu},\vec{B})=E_0+\sum_{ij} B_i \sigma_{ij}  \mu_j+ \text{higher order
terms}.
\end{equation}
Here $E_0$ is the electronic energy in the absence of nuclear moments and
external field. Note that the energy contains no terms of odd degree, which is
a consequence of the assumption that $E_0$ is a non-degenerate eigenvalue of a
time-even Hamiltonian. The second term in \eqref{energyexp} has
the form of a Zeeman interaction between the nuclear moment $\bm{\mu}$ and the
induced field $-\vec{B}\cdot\bm{\sigma}$. The $3\times 3$ matrix $\bm{\sigma}$
is the magnetic shielding tensor of the nucleus. 
We consider $\bm{\sigma}$ to be field independent, but
note that field-dependent corrections to the shielding can be found among the
higher order terms of \eqref{energyexp}. It follows from \eqnref{energyexp}
that $\sigma_{ij}$ can be obtained from perturbation theory (up to second
order) on $E_0$. Thus, in Ramsey's theory,
\begin{equation*}
\sigma_{ij}=\left.\frac{\partial^2E}{\partial B_i\partial\mu_j}\right\vert_0
\end{equation*}

For reasons of simplicity we have assumed that there is only one magnetic
nucleus in the molecule. When there is more than one, \eqnref{energyexp} is
readily adjusted to expand the energy $E(\bm{\mu}_1,\bm{\mu}_2,\ldots,\vec{B})$
in all the nuclear moments. Each nucleus will have its own shielding
tensor.

\section{Arbitrary electronic spectrum: Shieldings as free energy derivatives}

\subsection{Separation of slow and fast dynamical variables: Free energy as an 
effective Hamiltonian for nuclear spins}

We have seen that in the case of a non-degenerate isolated ground state the
shielding tensor is obtained from the electronic energy. In this section we
want to extend the theory to include molecules which have several
thermally populated electronic energy levels, so that temperature comes into
play. We want to find an expression for the shielding tensor that is
universally valid, regardless of the particular electronic spectrum of the
molecule. We start again from the physical assumptions of Ramsey: the nuclear
magnetic moments are classical vectors, they interact with the electrons, and
the electron cloud is at all instants in a state of equilibrium governed by the
applied magnetic field and the instantaneous orientation of the nuclear
moments. This equilibrium state is not simply the ground state, because we have
to allow for the molecule to possess multiple electronic states between which
rapid transitions occur. As the timescale of these transitions is much
shorter than the timescale of the nuclear spin dynamics,\cite{Kurland1970,
Gunne2009} the electron system can at all times preserve \emph{thermal}
equilibrium at constant temperature $T\equiv 1/\beta$. That is, the state of
the system is described by a Boltzmann distribution over the energy levels
$E_n(\bm{\mu},\vec{B})$, eigenvalues of the electronic Hamiltonian
$H(\bm{\mu},\vec{B})$. As the energy levels depend on $\bm{\mu}$, so do the
partition function and all other thermodynamic functions. Note that $\bm{\mu}$
is not treated as a dynamic variable of the thermodynamic system (which
consists of the electrons only) but as an external parameter that modifies the
energy levels of the system. 

Our next task is to establish which equation should replace \eqnref{energyexp}.
Recall that \eqnref{energyexp} gives the electronic energy as a function of
nuclear moments and external field. For the present purpose however, a better
interpretation is to regard this energy as the effective Hamiltonian
\cite{Kiang1992} for the system of nuclear moments in the external field
$\vec{B}$. There is no longer any explicit reference to the electrons but the
influence of the electrons on the motion of the nuclear moments enters the
effective Hamiltonian in such quantities as the shielding tensor, the nuclear
spin-spin coupling, etc.  For the class of molecules to which
\eqnref{energyexp} applies (non-degenerate, isolated ground state) we can thus
write: $H_\text{eff}(\bm{\mu};\vec{B}) = E(\bm{\mu},\vec{B})$. Note that in the
effective Hamiltonian, $\bm{\mu}$ is a dynamical variable, whereas $\vec{B}$
remains an external parameter (hence the semicolon). In the more general case
of electrons in thermal equilibrium the effective Hamiltonian is
temperature-dependent and is given\cite{Kiang1992} by the electronic Helmholtz
free energy $F=U-TS$, where $F$, $U$, and $S$ are parametric functions of
$\bm{\mu}$ and $\vec{B}$, and of temperature: 
\begin{equation}\label{Fexp}
H_\text{eff}(\bm{\mu};\vec{B}) = F(\bm{\mu},\vec{B}) = F_0+ \sum_{ij} B_i
\sigma_{ij}  \mu_j+ \text{higher order terms}.
\end{equation}
Here we have expanded $F$ in a Taylor series in analogy with the expansion of
the ground state energy in \eqnref{energyexp}. 
One way to see that $F$ acts as the effective
Hamiltonian for the nuclear moments is by considering the work done on the
system by changing an external parameter ($\mu_i$, say) at constant
temperature. This amount of work is given by the concomitant negative change in
$H_\text{eff}$. The only thermodynamic function that has this property is the
Helmholtz free energy. Hence, up to an irrelevant constant, the effective
Hamiltonian is given by $F$. 

\subsection{Analytical derivatives of the Helmholtz free energy}

The shielding tensor component  $\sigma_{ij}$ is the coefficient of $B_i\mu_j$
in the effective nuclear Hamiltonian. In Ramsey's theory $\sigma_{ij}$ is
temperature independent and can be obtained from Rayleigh--Schr\"odinger
perturbation theory on the ground state energy, \eqnref{energyexp}.  In the
general theory $\sigma_{ij}$ is temperature dependent and can be obtained from
perturbation theory on the free energy. From \eqnref{Fexp} we have
\begin{equation}\label{sigma_F}
\sigma_{ij}=\left.\frac{\partial^2F}{\partial B_i\partial\mu_j}\right\vert_0.
\end{equation}

We write the electronic Hamiltonian
$H(\bm{\mu},\vec{B})=H_0+V(\bm{\mu},\vec{B})$. Here $H_0$ is the electronic
Born--Oppenheimer Hamiltonian in the absence of nuclear magnetic moments and
external magnetic field, and $V(\bm{\mu},\vec{B})$ collects all interactions of
the electrons with $\bm{\mu}$ and $\vec{B}$. The expressions of these terms are
well known and can be found for example in Abragam.\cite{Abragam1961} Their
detailed form is not important for our discussion. 

In the calculation of $\sigma_{ij}$, only those terms of $V$ will contribute
that are linear in $\bm{\mu}$ or linear in $\vec{B}$ or bilinear in $\bm{\mu}$
and $\vec{B}$. Thus we write $V$ as the sum of four parts:
\begin{equation}\label{magnpert}
\begin{split}
V_\text{z} & = - \vec{m}\cdot \vec{B} = -\sum_i m_i B_i,\\
V_\text{hf} & = \bm{\mathcal{F}}\cdot \bm{\mu} = \sum_i \mathcal{F}_i\mu_i, \\
V_\mathcal{D} & = \vec{B}\cdot \bm{\mathcal{D}} \cdot \bm{\mu} = \sum_{ij}
B_i\mathcal{D}_{ij} \mu_j,\\
V' & = \text{terms of higher order in } \vec{B} \text{ and } \bm{\mu}. 
\end{split}
\end{equation}
Here $V_\text{z}$ is the electronic Zeeman Hamiltonian, with
$\vec{m}=-(\vec{L}+g_e\vec{S})$; $V_\text{hf}$ is the hyperfine coupling, and
$\bm{\mathcal{F}}$ may be further divided into orbit, spin dipole, and Fermi
contact terms;\cite{Abragam1961} $V_\mathcal{D}$ is the diamagnetic nuclear--magnetic
field coupling; $V'$ contains all those terms that are not needed to
calculate the shielding tensor, and will be discarded from now on. In general,
$V_\text{hf}$ and $V_\mathcal{D}$ must be summed over all magnetic nuclei, each
with its own $\bm{\mathcal{F}}$ and $\bm{\mathcal{D}}$, as these depend on the
position of the nucleus which they represent. It is important to keep in mind
that $\vec{m}$, $\bm{\mathcal{F}}$, and $\bm{\mathcal{D}}$ are electron
operators, whereas $\bm{\mu}$ and $\vec{B}$ are classical external parameters.
Furthermore, $\vec{m}$ and $\bm{\mathcal{F}}$ are time-odd, whereas
$\bm{\mathcal{D}}$ is time-even.  

For the derivation\cite{Feynman1972} of a perturbation expansion of $F$ it
will be convenient to introduce a parameter $\lambda$ in the Hamiltonian to
denote the combined order in $\bm{\mu}$ and \vec{B}:
\begin{equation}\label{H_lambda}
H=H_0+\lambda V_1+ \lambda^2V_2,
\quad \text{with}\quad V_1=V_\text{z}+V_\text{hf} \text{ and }
V_2=V_\mathcal{D}
\end{equation}
The free energy is given by
\begin{equation*}
F=-\frac{1}{\beta}\ln\tr\rho, \qquad \text{where}\quad\rho=e^{-\beta H}
\end{equation*}
Now $\rho$ can be expressed as a power series in $\lambda$:
\begin{equation}\label{rhoexpand}
\rho = \rho_0+\lambda \rho_1 + \lambda^2 \rho_2 +\ldots.
\end{equation}
Here $\rho_0=e^{-\beta H_0}$ and $\rho_1$ and $\rho_2$ are first and second
order corrections. A power series for $F$ can now be obtained as follows:
\begin{equation*}
\begin{split}
F&=-\frac{1}{\beta}\ln(\tr\rho_0+\lambda\tr\rho_1+ \lambda^2\tr\rho_2+\ldots)\\
 &=-\frac{1}{\beta}\ln\tr\rho_0-\frac{1}{\beta}
 \ln\left(1+\lambda\frac{\tr\rho_1}{\tr\rho_0}+\lambda^2\frac{\tr\rho_2}{\tr\rho_0}
 +\ldots\right)\\
 &=F_0 -\frac{\lambda}{\beta}\frac{\tr\rho_1}{\tr\rho_0}+
 \frac{\lambda^2}{\beta}
 \left[\frac{1}{2}\left(\frac{\tr\rho_1}{\tr\rho_0}\right)^2-
 \frac{\tr\rho_2}{\tr\rho_0}\right]+\ldots.
\end{split}
\end{equation*}
We know however that the free energy is an even function of the time-odd fields
\vec{B} and $\bm{\mu}$ and therefore odd powers of $\lambda$ must not occur in
the expansion of $F$. Hence $\tr\rho_1=0$ and
\begin{equation}\label{F_lambda2}
F=F_0-\frac{\lambda^2}{\beta}\frac{\tr\rho_2}{\tr\rho_0}+\ldots.
\end{equation}
To proceed, an expression for $\rho_2$ is needed.  $\rho$ satisfies the
Schr\"odinger equation $\frac{\partial\rho}{\partial\beta}=-H\rho$. By
substituting Eqs.\ \eqref{H_lambda} and \eqref{rhoexpand} and collecting terms
of equal power in $\lambda$, a recurrence relation for $\rho_i$ is found:
\begin{equation*}\label{rhorecurr}
\begin{split}
\frac{\partial\rho_1}{\partial\beta}&=-H_0\rho_1-V_1\rho_0,\\
\frac{\partial\rho_2}{\partial\beta}&=-H_0\rho_2-V_1\rho_1-V_2\rho_0,
\end{split}
\end{equation*}
etc. These are linear differential equations of the form
$\frac{dy}{dx}=ay+b(x)$, whose general solutions are given by
$y=e^{ax}\int_0^xe^{-at}b(t)dt$. Thus we find
\begin{equation*}
\begin{split}
\rho_1&=-\int_0^\beta e^{(w-\beta) H_0}V_1e^{-w H_0}dw\\
\rho_2&=-\int_0^\beta e^{(w-\beta) H_0}V_2e^{-w H_0}dw+
\int_0^\beta \int_0^w e^{(w-\beta) H_0}V_1e^{(w'-w)H_0}V_1e^{-w'\!H_0}
dw'\,dw.
\end{split}
\end{equation*}
Taking the trace of $\rho_2$, we use $\tr(AB)=\tr(BA)$ to rearrange and simplify 
the integrands.\cite{Feynman1972} Write $\rho_2=\rho_2^{(1)}+\rho_2^{(2)}$, then
\begin{equation*}
\tr\rho_2^{(1)}=-\int_0^\beta\tr[e^{-\beta H_0}V_2]dw=-\beta\langle
V_2\rangle_0\tr\rho_0,
\end{equation*}
where
\begin{equation*}
\langle V_2\rangle_0=\frac{\tr (V_2\rho_0)}{\tr\rho_0}
\end{equation*}
is the thermal average of $V_2$ in the canonical ensemble corresponding to
$H_0$.  
For the second part of $\rho_2$ we have
\begin{equation*}
\tr\rho_2^{(2)}=\int_0^\beta \int_0^w \tr[e^{-\beta H_0}e^{(w-w')H_0} 
 V_1e^{(w'-w)H_0}V_1]dw'\,dw.
\end{equation*}
The integrand depends on $w-w'$ only, suggesting a change of
variables $u=w$, $v=w-w'$. This gives
\begin{equation*}
\tr\rho_2^{(2)}= \int_0^\beta \int_0^u \tr[e^{-\beta H_0}e^{vH_0}V_1e^{-vH_0}V_1]dv\,du.
\end{equation*}
A second change of variables $u\rightarrow\beta-u$, $v\rightarrow\beta-v$ shows
that this integral is also equal to $\int_0^\beta\int_u^\beta$ of the same
integrand. Hence 
\begin{equation*}
\begin{split}
\tr\rho_2^{(2)}&= \frac{1}{2} \int_0^\beta \int_0^\beta \tr[e^{-\beta
H_0}e^{vH_0}V_1e^{-vH_0}V_1]dv\,du\\
 &= \frac{\beta}{2}\left\langle\int_0^\beta
 e^{vH_0}V_1e^{-vH_0}V_1dv\right\rangle_0\tr\rho_0.
\end{split}
\end{equation*}
At this point $\lambda$ is no longer needed and we set $\lambda=1$,
so that \eqnref{F_lambda2} becomes
\begin{equation*}
F= F_0 + \langle V_2\rangle_0-\frac{1}{2}\left\langle\int_0^\beta
 e^{wH_0}V_1e^{-wH_0}V_1dw\right\rangle_0+\ldots.
\end{equation*}
We substitute $V_1$ and $V_2$ according to Eqs.\ \eqref{H_lambda} and
\eqref{magnpert} and then take the second derivative as in \eqnref{sigma_F} to
obtain the shielding
\begin{equation}\label{sigma1}
\sigma_{ij}=\left.\frac{\partial^2F}{\partial B_i\partial\mu_j}\right\vert_0
 = \langle \mathcal{D}_{ij}\rangle_0 + \left\langle\int_0^\beta
  e^{wH_0}m_ie^{-wH_0}\mathcal{F}_jdw\right\rangle_0.
\end{equation}
In deriving this expression we have used that, for any two operators $A$ and $B$,
\begin{equation*}
\left\langle\int_0^\beta e^{wH_0}Ae^{-wH_0}B \,dw\right\rangle_0 = 
\left\langle\int_0^\beta e^{wH_0}Be^{-wH_0}A\, dw\right\rangle_0,
\end{equation*}
which can be shown by a change of integration variable $w\rightarrow \beta -w$. 

To evaluate the integral in \eqnref{sigma1}, the ensemble averaging is carried out
before the integration. Let $\ket{n\,\nu}$ be the eigenstates of $H_0$ with
eigenvalue $E_n(0,0)$, hereafter simply written $E_n$. The index $\nu$ labels
an arbitrary orthonormal basis of states with the same energy $E_n$. The
integral then becomes
\begin{multline*}
\int_0^\beta \left\langle e^{wH_0}m_ie^{-wH_0}\mathcal{F}_j\right\rangle_0 dw
 = \\ \frac{1}{Q_0}\sum_{n\nu,m\mu} \bra{n\,\nu}m_i\ket{m\,\mu} \bra{m\,\mu}
 \mathcal{F}_j \ket{n\,\nu}e^{-\beta E_n} \int_0^\beta e^{w(E_n-E_m)}dw,
\end{multline*}
where $Q_0 =\tr\rho_0=\sum_{n\nu}e^{-\beta E_n}$ denotes the partition function of
$H_0$. The remaining integral can now be evaluated and this gives
\begin{multline}\label{integral}
\int_0^\beta \left\langle e^{wH_0}m_ie^{-wH_0}\mathcal{F}_j\right\rangle_0 dw = 
\frac{1}{Q_0}\sum_{n}e^{-\beta E_n}\biggl[\beta \sum_{\nu,\nu'}\bra{n\,\nu}m_i\ket{n\,\nu'}
 \bra{n\,\nu'} \mathcal{F}_j\ket{n\,\nu}\\
+\sum_{m\neq n}\sum_{\nu,\mu} \frac{\bra{n\,\nu}m_i\ket{m\,\mu} 
 \bra{m\,\mu} \mathcal{F}_j\ket{n\,\nu} +\text{c.c.}}{E_m-E_n}\biggr].
\end{multline}

Finally, combining Eqs.\ \eqref{sigma1} and \eqref{integral}, we find a sum over
states formula for the nuclear shielding tensor,
\begin{multline}\label{sigma}
\sigma_{ij}=\frac{1}{Q_0} \sum_{n}e^{-\beta E_n}\biggl[\beta \sum_{\nu,\nu'}
\bra{n\,\nu}m_i\ket{n\,\nu'}\bra{n\,\nu'}\mathcal{F}_j\ket{n\,\nu}+\sum_\nu
\bra{n\,\nu}\mathcal{D}_{ij}\ket{n\,\nu}\\
+ \sum_{m\neq n}\sum_{\nu,\mu} \frac{\bra{n\,\nu}m_i\ket{m\,\mu} 
 \bra{m\,\mu} \mathcal{F}_j\ket{n\,\nu} +\text{c.c.}}{E_m-E_n}\biggr].
\end{multline}
Incidentally, the complex conjugate (c.c.) is not strictly needed and may be
replaced by a factor of 2 in front of the last sum, because this sum is always
real.  The present expression bears an evident similarity to the Van Vleck
equation for the magnetic susceptibility.\cite{Gerloch1975} One only has to
substitute $m_j$ for $\mathcal{F}_j$ and the diamagnetic susceptibility term
for $\mathcal{D}_{ij}$ to obtain the Van Vleck equation. The origin of this
similarity is readily understood when one notices that the susceptibility
tensor, just like the shielding tensor, occurs in a term of the free energy
expansion \eqnref{Fexp}, viz.\ in the term that is quadratic in the external
field: $\sum_{ij}B_i\chi_{ij}B_j$. The calculation of $\chi_{ij}$ then proceeds
along exactly identical lines and yields the Van Vleck equation.

\eqnref{sigma} can be cast in a more transparent form using projection
operators. Let $P_n=\sum_\nu \ket{n\,\nu}\bra{n\,\nu}$ be the projector on
level $n$ (which is possibly degenerate), and $Q_n=1-P_n$ its complement, then
we can write
\begin{multline*}
\sigma_{ij}=\frac{1}{Q_0} \sum_{n}e^{-\beta E_n} \tr 
 \biggl[\beta P_nm_iP_n\mathcal{F}_jP_n+P_n\mathcal{D}_{ij}P_n\\
 +P_nm_i\frac{Q_n}{H_0-E_n}\mathcal{F}_jP_n+P_n\mathcal{F}_j\frac{Q_n}{H_0-E_n}m_iP_n 
 \biggr].
\end{multline*}
This formula shows that the shielding tensor can be seen as a Boltzmann average
over the electronic energy levels of the unperturbed molecule, of a quantity
(the trace of the operator in brackets) that is associated with each level.
This quantity has a temperature-dependent part and a temperature-independent
part. The temperature-dependent part, 
\[
\beta\tr(P_nm_iP_n\mathcal{F}_jP_n), 
\]
which we shall call the Curie term, can
only be non-zero if $E_n$ is \emph{degenerate}. This is because if $E_n$ is not
degenerate, the eigenstate $\ket{n}$ is real (w.r.t.\ time-reversal) and
expectation values of time-odd operators on a real state vanish (i.e,
$P_nm_iP_n = P_n\mathcal{F}_jP_n = 0$ if $E_n$ is not degenerate).
Physically, the Curie term arises from the polarization in the applied field of
the permanent magnetic moment of the degenerate level. The
temperature-independent part of the quantity in brackets consists of two terms,
the `diamagnetic' term ($\sim\mathcal{D}$) and the `paramagnetic' term.  These
terms are known from the Ramsey theory for shielding in a non-degenerate ground
state. When the state is degenerate, the Curie term appears on top of the
Ramsey terms, shifting the chemical shift from the value it would have if the
state were not degenerate. This shift, coming from the Curie term, is known
as the paramagnetic shift. 

As \eqnref{sigma} was derived for a system with an arbitrary electronic
spectrum, it must obviously also be valid for a `closed shell' system, i.e.,
for a system with a non-degenerate, isolated ground state. In that case,
\eqnref{sigma} reduces to
\begin{equation}\label{sigma0}
\sigma_{ij}= \bra{0}\mathcal{D}_{ij}\ket{0}+\sum_{\substack{n\nu\\n \neq0}}
\frac{\bra{0}m_i\ket{n\,\nu}  \bra{n\,\nu} \mathcal{F}_j\ket{0}
+\text{c.c.}}{E_n-E_0}
\end{equation}
which is indeed the Ramsey expression, which can also be obtained starting from
\eqnref{energyexp} and applying perturbation theory (first order in
$\bm{\mathcal{D}}$ and second order in $\bm{\mathcal{F}}$ and $\vec{m}$) on the
ground state energy. The equivalence of the two approaches, one starting from
the ground state energy $E$, the other from the free energy $F$, 
follows from the fact that the entropy is zero (as only the ground state is
occupied), and therefore $F=E$. We would like to note that although
\eqnref{sigma0} has the same form as the original Ramsey expression,
\cite{Ramsey1950} it is more general than the latter in that the ground state
$\ket{0}$ in Ramsey's paper was assumed to be a pure spin singlet state, which
allowed him to ignore the spin-dependent parts of $\bm{\mathcal{F}}$ and
$\vec{m}$. The present treatment imposes no restrictions on $\ket{0}$ other
than those stated before, viz.\ that $\ket{0}$ is the non-degenerate ground
state of a time-even Hamiltonian $H_0$ (which, incidentally, excludes odd
electron systems). $H_0$ may include spin-orbit coupling and in cases where
this is an important effect the spin-dependent terms, the spin Zeeman term in
$\vec{m}$ and the spin-dipolar and Fermi contact hyperfine terms in
$\bm{\mathcal{F}}$, must be included in the calculation of $\sigma_{ij}$ from
\eqnref{sigma0}.

It should be noted that the procedure outlined above for the shielding tensor
can be used to obtain expressions for any desired term in the nuclear effective
Hamiltonian. It is, for example, straightforward to obtain nuclear spin-spin
coupling tensors by using the appropriate $V$ terms in the Hamiltonian and
carrying out the subsequent derivation as before. Thus, in analogy with
\eqnref{sigma1}, the coupling tensor between two nuclei $K$ and $L$ is found to
be
\begin{equation*}
J_{ij}^{KL}=\left.\frac{\partial^2F}{\partial \mu^K_i\partial\mu^L_j}\right\vert_0
 = \langle \mathcal{D}^{KL}_{ij}\rangle_0 + \left\langle\int_0^\beta
   e^{wH_0}\mathcal{F}_i^Ke^{-wH_0}\mathcal{F}_j^Ldw\right\rangle_0,
\end{equation*}
where $\mathcal{D}^{KL}_{ij}$ is the diamagnetic nucleus-nucleus coupling
term.\cite{Abragam1961}

\section{Shielding tensor in a degenerate isolated ground state}

We now consider a case of particular interest: a system with a degenerate
ground state of multiplicity $\omega$. We assume that the system is in thermal
equilibrium at a temperature such that no excited states are thermally
populated. The main results of this section  were published in Ref.\
\onlinecite{VandenHeuvel2012}. Here and in the appendices we provide a 
thorough discussion of the details of their derivation.

Applying our general equation \eqref{sigma} to the case of a degenerate and
isolated ground state we find
\begin{multline*}
\sigma_{ij}=\frac{1}{\omega}\biggl[\beta\sum_{\nu,\nu'}\bra{0\,\nu}m_i\ket{0\,\nu'}
\bra{0\,\nu'}\mathcal{F}_j\ket{0\,\nu}\\+ \sum_\nu
\bra{0\,\nu}\mathcal{D}_{ij}\ket{0\,\nu}
+ \sum_{m\neq0}\sum_{\nu,\mu}\frac{\bra{0\,\nu}m_i\ket{m\,\mu}
\bra{m\,\mu}\mathcal{F}_{j}\ket{0\,\nu}+\text{c.c.}}{E_m-E_0} \biggl].
\end{multline*}
The last two terms are familiar from the Ramsey expression \eqnref{sigma0} (and
reduce to the latter when $\omega=1$), but here they are averaged over the
states of the manifold. The first term is the Curie term which, as we have
noted before, is unique to degenerate states and causes a paramagnetic shift,
inversely proportional to temperature. As much of the following discussion will
focus on this term we repeat it explicitly:
\begin{equation}\label{parashift}
\sigma_{ij}^\text{p}=\frac{\beta}{\omega}\sum_{\nu,\nu'}\bra{0\,\nu}m_i\ket{0\,\nu'}
\bra{0\,\nu'}\mathcal{F}_j\ket{0\,\nu}.
\end{equation}

The sum in \eqnref{parashift} is evidently the trace of the product of two
matrices, respectively the representations of the electronic magnetic moment
and the hyperfine field on the nucleus in a basis of the degenerate manifold.
Being a trace of a matrix product, it is invariant under unitary
transformations of the basis, which is of course required for an observable
property. This means that we are free to choose whichever orthonormal
basis we want to calculate $\sigma_{ij}^\text{p}$ using \eqnref{parashift}. 

Both matrices in \eqnref{parashift} are known from the theory of electron
paramagnetic resonance (EPR) spectroscopy. They represent respectively the
Zeeman effect and the hyperfine coupling in the degenerate manifold. And thus
we arrive at the well-known result that the paramagnetic shielding tensor
can be obtained completely from EPR parameters, and, vice versa, NMR experiments
on paramagnetic molecules provide information on the EPR parameters. 
The precise expression of this correspondence will be derived in section 
\ref{sec:arbmult}, but let us start with an example, the simplest case of degeneracy, a Kramers doublet.

\subsection{Kramers doublet}\label{sec:KD}
In a Kramers doublet the Zeeman and hyperfine interactions can be conveniently
expressed by means of a spin Hamiltonian 
\begin{equation}\label{spinhKD}
H_S=\bohr \vec{S}\cdot\mat{g}\cdot\vec{B}+\vec{S}\cdot\mat{A}\cdot\vec{I}
\end{equation}
operating within a fictitious spin-1/2 doublet.\cite{Abragam_EPR} What this
means is that the matrix elements of $V_z+V_\text{hf}$ (\eqnref{magnpert})
within the true wave functions of the Kramers doublet are the same as the
matrix elements of $H_S$ within the fictitious spin
doublet.\cite{Griffith1960,Griffith1967} For example, if we denote the basis
states of the Kramers doublet by $\ket{0\,a}$ and $\ket{0\,b}$ and let
$\ket{0\,a}$ correspond to the fictitious spin state $\ket{\frac{1}{2}}$ and
$\ket{0\,b}$ to $\ket{-\frac{1}{2}}$, then we have
\begin{equation*}
\begin{split}
\bra{0\,a}m_x\ket{0\,b}&=
-\bohr\braket{\tfrac{1}{2}}{{\textstyle\sum_i} S_ig_{ix}}{-\tfrac{1}{2}}=
-\bohr(g_{xx}-ig_{yx})/2,\\
\bra{0\,a}\mathcal{F}_x\ket{0\,b}&= \frac{1}{g_I\nbohr}
\braket{\tfrac{1}{2}}{{\textstyle\sum_i}S_iA_{ix}}{-\tfrac{1}{2}}=
\frac{1}{g_I\nbohr} (A_{xx}-iA_{yx})/2,
\end{split}
\end{equation*}
and so on. Here $g_I$ is the nuclear $g$-factor and we have used 
$\bm{\mu}=g_I\nbohr \vec{I}$, where $\vec{I}$ is the nuclear spin vector in
units of $\hbar$. Using the spin Hamiltonian of the Kramers doublet,
\eqnref{parashift} becomes 
\begin{equation*}
\sigma_{ij}^\text{p}= -\frac{\bohr \beta}{2g_I\nbohr} \sum_{k,l} g_{ki}A_{lj}
\tr(S_kS_l).
\end{equation*}
The trace of the product of two cartesian spin components is given by
\begin{equation*}
\tr(S_kS_l) = \delta_{kl} \frac{S(S+1)(2S+1)}{3}.
\end{equation*}
And with $S=1/2$, we find that the paramagnetic shielding tensor for a Kramers
doublet can be expressed as follows.
\begin{equation}\label{KDtensor}
\bm{\sigma}^\text{p} = -\frac{\bohr \beta}{4g_I\nbohr}\, \mat{g}^
\transp\mat{A}.
\end{equation}
This equation was first obtained by Moon and Patchkovskii using a
different method.\cite{Moon2004} Note how the two interactions that combine
to give rise to a paramagnetic shift are separated in \eqnref{KDtensor}:
$\mat{g}$ represents the Zeeman splitting in the external field and is a
property of the molecule as a whole, independent of the nuclear moments;
$\mat{A}$ represents the hyperfine interaction between the electrons and the
nucleus whose shielding tensor we wish to calculate. The ratio $\mat{A}/g_I$
however is independent of the nuclear species and depends only on the position
of the nucleus in the molecule.

\subsection{Degeneracy of arbitrary multiplicity}\label{sec:arbmult}

The example of the Kramers doublet showed how the paramagnetic shielding can be
obtained from EPR $g$- and $A$-tensors. In this section we extend this
treatment to degenerate ground states of any multiplicity and derive a formula
for $\bm{\sigma}^\text{p}$ which is the generalization of
\eqnref{KDtensor}.\cite{VandenHeuvel2012}

The starting point is again \eqnref{parashift}. We have seen that the matrices
of $m_i$ and $\mathcal{F}_i$ in a Kramers doublet can be reproduced by 
operator equivalents, linear in the spin components, working in a fictitious spin
doublet. An expansion of this kind is an application of a general theorem to the
effect that \emph{any} $n\times n$ matrix can be reproduced by a \emph{unique}
spin operator, polynomial in the spin components $S_i$.\cite{Griffith1960}
Mathematical details about the construction and properties of this spin
operator are given in Appendix \ref{app:spinop}. The main result is that the
matrix of an operator $X$ in a basis $\{\psi_i\}_{i=1\ldots n}$ can be
reproduced by a spin operator $X_S$ in a \emph{fictitious} spin multiplet
$\{\ket{S\,M}\}$ whose dimension equals $n$, i.e., $2S+1=n$.
Specifically,\cite{Griffith1960}
\begin{equation*}
X_S = \sum_{k=0}^{2S}\sum_{q=-k}^k (-1)^q Q_{q}^{(k)} \ito{k}{-q}.
\end{equation*}
where $Q_{q}^{(k)}$ are complex coefficients, unique for given operator
$X$ and basis functions $\psi_i$. The \ito{k}{q} are irreducible tensor
operators, which are basically products of $k$ spin components $S_i$ adapted to
spherical symmetry.\cite{Zare1988} The formula for the paramagnetic shielding,
\eqnref{parashift}, depends on six matrices, one for each cartesian component
of the magnetic moment $\vec{m}$ and the hyperfine field $\bm{\mathcal{F}}$.
We now apply the spin-decomposition to each of these matrices and write the
corresponding spin operators
\begin{equation}\label{spinopmF}
\begin{split}
m_{i,S}&=-\bohr \sum_{k=0}^{2S}\sum_{q=-k}^k (-1)^q \ito{k}{-q} g^{(k)}_{qi}\\
\mathcal{F}_{i,S}&= \frac{1}{g_I\nbohr} \sum_{k=0}^{2S}\sum_{q=-k}^k (-1)^q
\ito{k}{-q} A^{(k)}_{qi}
\end{split}
\end{equation}
where the $Q$-numbers have been renamed $g^{(k)}_{qi}$ for the magnetic moment
and $A^{(k)}_{qi}$ for the hyperfine field, in accordance with the usual
notation in EPR spectroscopy. Note that in the present formalism the
``$g$-factors'' and ``$A$-factors'' are complex numbers whose complex
conjugates are given by 
\begin{equation*}
{g^{(k)}_{qi}}^*=(-1)^qg^{(k)}_{-qi},\quad
{A^{(k)}_{qi}}^*=(-1)^qA^{(k)}_{-qi}.
\end{equation*}
These relations make the operators in \eqnref{spinopmF} Hermitian (See
Appendix, \eqnref{spinopherm}).
The spin Hamiltonian 
\begin{equation*}
\begin{split}
H_S&=-\sum_i m_{i,S}B_i + g_I\nbohr\sum_i\mathcal{F}_{i,S}I_i\\
   &= \bohr\sum_i\sum_{kq} (-1)^q \ito{k}{-q} g^{(k)}_{qi}B_i +
     \sum_i\sum_{kq} (-1)^q\ito{k}{-q} A^{(k)}_{qi}I_i
\end{split}
\end{equation*}
gives a complete description of an EPR experiment in the degenerate manifold.
It is the generalization of \eqnref{spinhKD} to a  manifold of arbitrary
degeneracy. 

We can now proceed to express the paramagnetic shielding tensor in terms of the
EPR parameters. In \eqnref{parashift}, we replace the matrix elements in real
space by the corresponding, and identical, matrix elements in fictitious spin
space:
\begin{equation*}
\begin{split}
\sigma_{ij}^\text{p}&=\frac{\beta}{\omega}\sum_{M,M'}
\braket{S\,M}{m_{i,S}}{S\,M'} \braket{S\,M'}{\mathcal{F}_{j,S}}{S\,M}\\
&= \frac{\beta}{2S+1} \tr (m_{i,S}\mathcal{F}_{j,S})\\
&=-\frac{\bohr}{g_I\nbohr} \frac{\beta}{2S+1} \sum_{kq}\sum_{k'q'}
  (-1)^{(q+q')}g^{(k)}_{qi} A^{(k')}_{q'j}
  \tr\left(\ito{k}{-q}\ito{k'}{-q'}\right)
\end{split}
\end{equation*}
The trace simplifies using Eqs.\ \eqref{hermkram} and \eqref{ito-ortho} and we
get
\begin{equation}\label{sigmageneral}
\sigma_{ij}^\text{p}=-\frac{\bohr}{g_I\nbohr} \frac{\beta}{2S+1} 
 \sum_{k=0}^{2S}\sum_{q=-k}^k {g^{(k)}_{qi}}^* A^{(k)}_{qj} 
 \frac{\abs{\bra{S}|\ito{k}{}|\ket{S}}^2}{2k+1}.
\end{equation}
This expression is the generalization of \eqnref{KDtensor}. Note that,
whereas the spin Hamiltonian parameters are basis
dependent (see Appendix \ref{app:spinop}), this is not so for the contraction of
the parameters appearing in \eqnref{sigmageneral}. The tensor
$\bm{\sigma}^\text{p}$ is thus an invariant property of the degenerate
manifold, as expected. Another such invariant is the magnetic susceptibility
tensor, which is proportional to the same contraction but with $g^{(k)}_{qj}$
substituted for $A^{(k)}_{qj}$. For the twofold degenerate Kramers doublet this
particular contraction is in fact well known as the \mat{G} tensor:
$\mat{G}=\mat{g}^\transp\mat{g}$, with \mat{g} as in
\eqnref{spinhKD}.\cite{Abragam_EPR} 

An example of the application of \eqnref{sigmageneral} to the cubic quartet
electronic ground state of certain paramagnetic lanthanide ion impurities in
CaF$_2$ crystals can be found in Ref.\ \onlinecite{VandenHeuvel2012}. In the
next section we apply the formula to the icosahedral sextet that can arise in
endohedral metallofullerene Ln@C$_{60}$.

\subsubsection{Example: paramagnetic shielding in the icosahedral sextet}

The highest degeneracy that can be generated by point group symmetry is sixfold
and corresponds to the W$'$ irreducible spin representation of the icosahedral
double group $I^*$.\cite{Griffith_TTMI} A ground state of this symmetry type
may arise from the crystal field splitting of the $J$ ground multiplet of the
rare earths Ce$^{3+}$, Nd$^{3+}$, Sm$^{3+}$, Dy$^{3+}$, Er$^{3+}$, and
Yb$^{3+}$, when placed in an environment of icosahedral
symmetry.\cite{Walter1987} We consider the paramagnetic shielding at the
central nucleus. 

The high symmetry reduces the number of free parameters in the spin Hamiltonian
drastically. To make maximum use of the symmetry it is best to have the
cartesian axes coincide with the twofold axes of a $D_2$ subgroup of $I$. This
way the $x$, $y$, and $z$ directions are equivalent,\cite{Qiu2002} and the
shielding tensor is isotropic and diagonal: $\bm{\sigma}^\text{p}=\sigma
^\text{p} \mat{I}$. The fictitious spin representation of W$'$ has $S=5/2$. A
vector operator equivalent working in this space has to transform as the T$_1$
irreducible representation of the icosahedral group.\cite{Griffith1960} This
means that the linear combinations of tensor operators in \eqnref{spinopmF}
must transform as the cartesian component of T$_1$. Symmetry
adaptations of spherical harmonics were published by Qiu and
Ceulemans.\cite{Qiu2002} Using their tables one finds that T$_1$ occurs in the
subduction of $k=1$ and $k=5$. The spin operators with $k=1$ are just the
cartesian spin
components $S_x$, $S_y$, and $S_z$. The corresponding operators for $k=5$ are
as follows (Ref.\ \onlinecite{Qiu2002}, Table 5. $\beta_{n,m}=\frac{1}{2}(\pm
n+m\sqrt{5})$)
\begin{equation*}
\begin{split}
T_x&=\tfrac{\sqrt{7}\beta_{1,3}}{16\sqrt{2}}\left(\ito{5}{5}-\ito{5}{-5}\right) 
     - \tfrac{3\sqrt{7}\beta_{-1,1}}{16\sqrt{2}}\left(\ito{5}{3}-\ito{5}{-3}\right)
          -\tfrac{\sqrt{3}\beta_{7,1}}{16}\left(\ito{5}{1}-\ito{5}{-1}\right)\\
T_y&=i\tfrac{\sqrt{7}\beta_{-1,3}}{16\sqrt{2}}\left(\ito{5}{5}+\ito{5}{-5}\right) 
     - i\tfrac{3\sqrt{7}\beta_{1,1}}{16\sqrt{2}}\left(\ito{5}{3}+\ito{5}{-3}\right)
               +i\tfrac{\sqrt{3}\beta_{-7,1}}{16}\left(\ito{5}{1}+\ito{5}{-1}\right)\\
T_z&=\tfrac{\sqrt{35}}{16}\left(\ito{5}{4}+\ito{5}{-4}\right)  
      +\tfrac{\sqrt{21}}{8}\left(\ito{5}{2}+\ito{5}{-2}\right)
      -\tfrac{3}{8\sqrt{2}}\ito{5}{0}
\end{split}
\end{equation*}
These expressions are normalized in the sense that the squares of the
coefficients sum up to one.
It remains now to choose a reduced matrix element for $S^{(5)}$.
\eqnref{sigmageneral} adopts a simplified form if we choose
\begin{equation*}
\frac{\abs{\bra{S}|\ito{5}{}|\ket{S}}^2}{11}=
\frac{\abs{\bra{S}|\ito{1}{}|\ket{S}}^2}{3}=
\frac{(2S+1)S(S+1)}{3}=\frac{35}{2},
\end{equation*}
for $S=5/2$. The second equality follows from our choice of $S^{(1)}$ as in
\eqnref{ito1}.\cite{Zare1988}

A completely general spin Hamiltonian for Zeeman and hyperfine coupling (with
the central nucleus) in the icosahedral sextet thus contains just four parameters
and is given by
\begin{equation*}
H_S= \bohr (g\vec{S}+g'\vec{T})\cdot \vec{B} + (A\vec{S}+A'\vec{T})\cdot
\vec{I}.
\end{equation*}
The expression for the paramagnetic shielding follows now readily from
\eqnref{sigmageneral}:
\begin{equation*}
\sigma^p=-\frac{35}{12}\frac{\bohr}{g_I\nbohr}\beta\bigl(gA+g'A'\bigr).
\end{equation*}


\begin{acknowledgments}
A.S. would like to acknowledge financial support from the Early Career
Researcher Grant (ECR 2012) scheme from the University of Melbourne, and from
the Interdisciplinary Seed Funding 2012 scheme from the Melbourne Materials
Institute and the University of Melbourne.
\end{acknowledgments}

\appendix*

\section{Spin operator equivalents}\label{app:spinop}

We consider the following problem. Given an arbitrary, complex square matrix of
dimension $2S+1$, to construct a spin operator, polynomial in the components of
$\vec{S}$, whose matrix representation in the fictitious spin manifold
$\ket{S\,M}$ is the same as the given matrix. The main results in this appendix
are not new and were published by
Griffith.\cite{Griffith1960,Griffith1963,Griffith1967} For convenience, we
collect them here in a coherent presentation, and expound in more detail on the
derivations.

\subsection{Orthogonal matrix expansion}
The first non-trivial case is that of a complex $2\times 2$ matrix. This
corresponds to fictitious spin $S=1/2$. It is not hard to see that any such
matrix can be expanded as 
\begin{equation*}
a \mat{I}+b\bm{\sigma}_x+c\bm{\sigma}_y+d\bm{\sigma}_z,
\end{equation*}
where $\mat{I}$ is the $2\times 2$ unit matrix and the $\bm{\sigma}_i$ are the
Pauli matrices:
\begin{equation*}
\bm{\sigma}_x=\begin{pmatrix}0&1\\1&0\end{pmatrix}, \quad
\bm{\sigma}_y=\begin{pmatrix}0&-i\\i&0\end{pmatrix}, \quad
\bm{\sigma}_z=\begin{pmatrix}1&0\\0&-1\end{pmatrix}.
\end{equation*}
The four complex quantities $a,b,c,d$ are uniquely determined by a set of four
linear equations, one for each element of the given matrix. The corresponding
spin operator is given by
\begin{equation}\label{spinop-2}
a+2bS_x+2cS_y+2dS_z.
\end{equation}
We have seen an example in section \ref{sec:KD}, where matrices of Zeeman and
hyperfine coupling in a Kramers doublet were represented by a spin Hamiltonian
(\eqnref{spinhKD}), which is an application of \eqnref{spinop-2}, (with
$a=0$, as the matrices considered there are traceless).

When we move to matrices of dimension $n\times n$ ($n\geq 3$), the operator
equivalent in \eqnref{spinop-2} is no longer adequate to reproduce an arbitrary
matrix. This is most easily seen from the requirement that the number of
parameters in the spin operator must be at least $n^2$ (i.e., the number of
elements in the arbitrary matrix), whereas there are only four of them in
\eqnref{spinop-2}. The most obvious way to increase the number of parameters is
to include terms of quadratic, cubic, and higher degrees in the spin components
$S_i$, until the required number of parameters is reached. This procedure
suffers from a drawback though, which is that linear dependencies arise among
the terms. For example, there are nine quadratic terms: $S_x^2$, $S_xS_y$,
$S_yS_x$, \ldots, but four linear combinations of them are already contained
among the terms of lower degree (i.e., in \eqnref{spinop-2}). These are
$S_x^2+S_y^2+S_z^2=\vec{S}^2=S(S+1)$, $S_xS_y-S_yS_x=iS_z$,
$S_yS_z-S_zS_y=iS_x$, and $S_zS_x-S_xS_z=iS_y$. The five remaining linear
combinations are linearly independent from terms of lower degree and can be
added to \eqnref{spinop-2} to obtain a spin operator with nine parameters in
total, which allows to expand any $3\times3$ matrix. 

This procedure may be repeated for higher degrees and one will always find
that, among the $3^k$ spin products of degree $k$, only $2k+1$ are 
independent from the products of lower degree. These $2k+1$ operators form an
irreducible tensorial set of rank $k$. To proof this we make use of the general
theory of irreducible tensor operators (ITO) and angular momentum, as can be
found for example in Zare.\cite{Zare1988}

Let \ito{k}{q} ($q=-k,\ldots,k$) denote the irreducible tensor operators
of rank $k$ constructed from the products of $k$ spin components $S_i$.
\ito{0}{0} is simply a constant, which may be taken to be 1. The three
components of \ito{1}{} are (up to a common, real factor of choice, see below)
\begin{equation}\label{ito1}
\ito{1}{0}=S_z, \qquad \ito{1}{\pm1}=\frac{1}{\sqrt{2}}(\mp S_x-iS_y).
\end{equation}
The tensors of higher rank are obtained as follows. \ito{2}{} is found by
coupling \ito{1}{} with itself into a rank-2 ITO. Next, \ito{3}{} is found by
coupling \ito{2}{} with \ito{1}{} into a rank-3 ITO, and so on. This process may be
repeated several times to obtain the ITO of desired rank. We thus define our
\ito{k}{q} by \eqnref{ito1} and the following recursion relation.
\begin{equation}\label{ito-recur}
\ito{k}{q}
= N_{k}(-1)^q\sum_{q_1,q_2}\tj{k-1}{1}{k}{q_1}{q_2}{-q}
\ito{k-1}{q_1}\ito{1}{q_2}.
\end{equation}
The multiplicative constant $N_k$ may be chosen freely for each $k$ but we
require it to be real, in order for the tensor operator to obey certain simple
relations under Hermitian and time conjugation, to be explained in Section
\ref{sec:conjugation} below.  The large bracket is a 3$j$ symbol.\cite{Zare1988}

We write \itom{k}{q} for the matrix representation of \ito{k}{q} in the basis
\ket{S\,M}. The Wigner--Eckart theorem tells us that the matrix elements
factor into a $3j$ symbol and a reduced matrix element as follows:
\begin{equation*}
\bra{S\,M}\ito{k}{q} \ket{S\,M'}=(-1)^{S-M} \bra{S}|\ito{k}{}|\ket{S}
 \tj{S}{k}{S}{-M}{q}{M'}.
\end{equation*}
Note that the triangle condition implies that the matrices are zero for all
$k>2S$. We can now show that the \itom{k}{q} ($k=0,1,\ldots,2S$) form
a basis for the vector space $\mathbb{C}^{n\times n}$ ($n=2S+1$) of all complex
$n\times n$ matrices. First, observe that the set consists of $\sum_{k=0}^{2S}
(2k+1) = (2S+1)^2=n^2$ matrices, in agreement with the dimension of
$\mathbb{C}^{n\times n}$. Second, the matrices constitute a linearly
independent set, which can be shown as follows. Define an inner product between
two matrices by imagining each matrix to be reorganized as a vector by sticking
the rows together and then taking the usual scalar product in
$\mathbb{C}^{n^2}$:
\begin{equation*}
(\mat{A},\mat{B})=\sum_{ij} A_{ij}^*
B_{ij}= \tr(\mat{A}^\dagger \mat{B}).
\end{equation*}
We have now
\begin{align}
 &\left({\itom{k}{q}},\itom{k'}{q'}\right)=
 \tr\left({\itom{k}{q}}^\dagger \itom{k'}{q'}\right)=
   \sum_{MM'} \bra{S\,M}\ito{k}{q}\ket{S\,M'}^*
   \bra{S\,M}\ito{k'}{q'} \ket{S\,M'} \notag\\
 &= \sum_{MM'}  \tj{S}{S}{k}{M'}{-M}{q}
   \tj{S}{S}{k'}{M'}{-M}{q'} \bra{S}|\ito {k}{}|\ket{S}^*
   \bra{S}|\ito{k'}{}|\ket{S} \notag\\
 &= \frac{\abs{\bra{S}|\ito{k}{}|\ket{S}}^2}{2k+1}\,
 \delta_{kk'}\delta_{qq'},\label{ito-ortho}
\end{align}
which shows that the \itom{k}{q} are orthogonal and therefore linearly
independent. Note that the derivation leading to \eqnref{ito-ortho} makes
nowhere use of the specific form one may choose for the \ito{k}{q}, such
as the one in \eqnref{ito-recur} for example. Rather, the result
follows straight from the Wigner--Eckart theorem, which is by definition true
for all ITO's.

It follows that any given $n\times n$ matrix \mat{X} has a unique expansion
in spin matrices \itom{k}{q} of the ITOs \ito{k}{q} in the basis \ket{S\,M}
($2S+1=n$). The expansion coefficients are obtained by orthogonal projection,
using \eqnref{ito-ortho}.
\begin{align}
\mat{X} &= \sum_{k=0}^{2S}\sum_{q=-k}^k (-1)^q Q_{q}^{(k)}
\itom{k}{-q}\label{matrixexp} \\
Q_{q}^{(k)} &= \alpha(k)(-1)^{q}\tr\left({\itom{k}{-q}}^\dagger
\mat{X}\right), \qquad \alpha(k)=\frac{2k+1}{\abs{\bra{S}|\ito{k}{}|\ket{S}}^2}
\nonumber
\end{align}

\subsection{The choice of basis}
We have established that to every square matrix \mat{X} there corresponds a
unique\footnote{Unique for a fixed choice of the \ito{k}{q}} set of
coefficients $Q_q^{(k)}$ such that \eqnref{matrixexp} is true. These
coefficients define the spin operator which reproduces \mat{X} in the fictitious
spin manifold: 
\begin{equation}\label{spinop}
X_S = \sum_{k=0}^{2S}\sum_{q=-k}^k (-1)^q Q_{q}^{(k)} \ito{k}{-q}.
\end{equation}
However, when \mat{X} is the matrix representation of a true operator $X$, the
spin equivalent $X_S$ representing $X$ is not at all unique. Given an
orthonormal basis $(\psi_1,\ldots,\psi_n)$ and
\begin{equation*}
X_{ij}=\braket{\psi_i}{X}{\psi_j},
\end{equation*}
any unitary transformation of this basis results in an equally valid
representation of $X$, given by a transformed matrix
$\mat{X}\rightarrow \mat{U}^\dagger\mat{X}\mat{U}$. The transformed
matrix may in turn be expanded according to \eqnref{matrixexp}, yielding a new set of
coefficients $Q_q^{(k)}$ and hence a new operator equivalent $X_S$. Clearly, an
infinite number of different spin operators can thus be generated and they all
represent the same physical situation. 
Choosing a basis can be seen as establishing an association between the true
states and the fictitious spin states: $\psi(M) \leftrightarrow \ket{S\,M}$. A
different association leads to a different spin operator, and a
suitable choice often leads to a simplified spin operator.\cite{Griffith1967} 

\subsection{Implications of Hermiticity and time
reversal}\label{sec:conjugation}
The matrices we wish to represent by spin operators are usually Hermitian
and they often have a symmetry related to time reversal as well. These two
properties will naturally impose some restrictions on the spin operator
equivalent. Under Hermitian conjugation ($\dagger$) and Kramers's time reversal
operation (*)\footnote{$A^*$ is shorthand for $KAK^\dagger$, where $K$ is the
time reversal operator (Ref.\ \onlinecite{messiah}, Chapter XV)} the \ito{k}{q}
transform as follows: 
\begin{equation}\label{hermkram}
\begin{split}
{\ito{k}{q}}^\dagger&= (-1)^q\ito{k}{-q}\\
{\ito{k}{q}}^*&= (-1)^{q-k}\ito{k}{-q},
\end{split}
\end{equation}
which can be verified from Eqs.\ \eqref{ito1} and \eqref{ito-recur}, using
induction on $k$.

\paragraph{Hermiticity.} Suppose now that \mat{X} is an Hermitian matrix, e.g., the matrix
representation of an Hermitian operator in an orthogonal basis of
wave functions. There exists a unique operator $X_S$, of the form
\eqref{spinop}, whose matrix in \ket{S\,M} equals \mat{X}. The Hermiticity of
\mat{X} then implies
\begin{equation*}
\braket{S\,M'}{X_S}{S\,M}= \braket{S\,M}{X_S}{S\,M'}^* = 
\braket{S\,M'}{X_S^\dagger}{S\,M},
\end{equation*}
for all $M,M'$, which can only be true if $X_S^\dagger=X_S$. Thus the
Hermiticity of an operator in true space translates into the Hermiticity of the
equivalent spin operator, and this is independent of the choice of basis in true
space. The fact that an Hermitian true operator must be represented by an
Hermitian spin operator may seem obvious, but we mention it explicitly because
the behavior of the true operator under time conjugation does \emph{not}
automatically lead to the same behavior of the spin operator, as we shall see
below. From
Eqs.\ \eqref{spinop} and \eqref{hermkram} follow the conditions on the
coefficients $Q_q^{(k)}$ that make $X_S$ Hermitian:
\begin{equation}\label{spinopherm}
X_S^\dagger=X_S \quad\Leftrightarrow \quad{Q_q^{(k)}}^*=(-1)^{q}Q_{-q}^{(k)}.
\end{equation}

\paragraph{Time reversal.}The operation of time reversal works on states and on
operators. We consider the implications hereof on the spin operator equivalent.
Let $X$ be the operator working in a set of $n$ states. We assume that this set
is closed under time reversal. This means that there exists a unitary matrix
\mat{A} such that
\begin{equation}\label{timeclosed}
\psi_i^*=\sum_{j=1}^n\psi_jA_{ji}.
\end{equation}
If $\psi_i$ describes a system of $n_e$ electrons then we know that
$(\psi_i^*)^*=(-1)^{n_e}\psi_i$, and therefore $\mat{A}\mat{A}^*=(-1)^{n_e}$.
From this and the fact that \mat{A} is unitary follows
$\mat{A}=(-1)^{n_e}\mat{A}^\transp$, that is, \mat{A} is symmetric for even and
antisymmetric for odd number of electrons. In spin space, the time reversal
operator is given by $K_S=Y_SK_0$, where $Y_S$ is the operator for a rotation
through $\pi$ about the $y$ axis and $K_0$ takes the complex conjugate of
numbers, but leaves the basis kets \ket{S\,M} unchanged.\cite{messiah}
\begin{equation}\label{spinTR}
\ket{S\,M}^*\equiv K\ket{S\,M}=Y_S\ket{S\,M}=(-1)^{S-M}\ket{S\,{-}M}.
\end{equation}

We shall consider the case where the operator $X$ has a definite parity under
time reversal: $X^*=\epsilon X$, with $\epsilon=\pm1$. Using
\eqnref{timeclosed} we have for the matrix elements of $X$
\begin{equation}\label{XTR}
X_{ij}^*=\braket{\psi_i}{X}{\psi_j}^* = \braket{\psi_i^*}{X^*}{\psi_j^*} =
\epsilon\sum_{k,l}A_{ki}^*X_{kl}A_{lj},
\end{equation}
or, in matrix notation, 
\begin{equation*}
\mat{X}^* = \epsilon \mat{A}^\dagger\mat{X}\mat{A}.
\end{equation*}
On the other hand, for the matrix elements of $X_S$ we find, using
\eqnref{spinTR}
\begin{equation}\label{STR}
\braket{S\,M'}{X_S}{S\,M}^* = \braket{S\,M'}{Y_S^\dagger X_S^*Y_S}{S\,M}
\end{equation}
We now define $A_S$ as the spin operator whose matrix representation in
\ket{S\,M} is \mat{A}. Then we can derive from Eqs.\ \eqref{XTR} and
\eqref{STR} and the requirement that $X_S$ reproduces the matrix \mat{X} that
\begin{equation*}
\braket{S\,M'}{Y_S^\dagger X_S^*Y_S}{S\,M} = \epsilon \braket{S\,M'}{A_S^\dagger
X_S A_S}{S\,M},
\end{equation*}
and therefore
\begin{equation}\label{spinopTR}
X_S^* = \epsilon \,Y_SA_S^\dagger X_S A_SY_S^\dagger.
\end{equation}
This result shows that, in general, the fact that the true operator $X$ has a
definite time parity $\epsilon$ does not automatically imply that the same is
true for the fictitious spin operator $X_S$. In particular, the behavior of
$X_S$ under time conjugation is seen to depend, through $A_S$, on the choice of
basis in the manifold of true states.

Is it nevertheless possible, by a suitable choice of basis, to obtain that
$X_S^*=\epsilon X_S$? This would require, from \eqnref{spinopTR}, that
$A_S=Y_S$. Then we would have 
\begin{equation}\label{TRbasistest}
(-1)^{n_e}=\mat{A}\mat{A}^*=\mat{Y}_S\mat{Y}_S^*=\mat{Y}_S^2=(-1)^{2S},
\end{equation}
where the third equality follows from the reality of the matrix elements of
$\mat{Y}_S$, see \eqnref{spinTR}, and the last equality follows from the fact that
$Y_S^2$ is a rotation through an angle of $2\pi$, which multiplies a spin ket
by $+1$ if $S$ is integer and by $-1$ if $S$ is half-integer. We conclude from
\eqnref{TRbasistest} that in cases where $n_e+2S$ is odd, we cannot in general
obtain that $X_S^*=\epsilon X_S$. 

Let us first then consider the other case, $n_e+2S$ even. Recall that the
fictitious spin quantum number $S$ is determined by the degeneracy of the
manifold $n=2S+1$. Thus, $n_e+2S$ is even for odd degeneracy in an even
electron system and for even degeneracy in an odd electron system. For these
cases it is always possible to find a basis $\psi(M)$ in true space satisfying
$\psi(M)^*= (-1)^{S-M}\psi(-M)$. With this choice we have $\mat{A}=\mat{Y}_S$
and thus a spin operator which has the same time parity as the true operator:
$X_S^*=\epsilon X_S$. If $X_S$ is also Hermitian we find, using Eqs.\
\eqref{spinop}, \eqref{hermkram} and \eqref{spinopherm},
\begin{equation}\label{Qrules1}
X_S^\dagger = X_S \text{ and } X_S^*=\epsilon X_S\quad \Leftrightarrow
\quad Q_q^{(k)}=0 \text{ for}
\begin{cases}
 \text{$\epsilon=+1$ and $k$ odd}\\
 \text{$\epsilon=-1$ and $k$ even}
\end{cases}
\end{equation}
This result is equivalent to the well-known rule that a spin Hamiltonian can
contain only even powers of the spin if it represents a time-even operator and
only odd powers of the spin if it represents a time-odd operator. However, this
rule is only valid for a particular choice of basis, namely one which
transforms under time reversal in exactly the same way as the associated spin
basis.

It remains now to discuss the case $n_e+2S$ odd. Although this case is far less
common than the previous one, it is worth including because the rule mentioned
in the previous paragraph does not apply here. First note that we may limit the
discussion to even electron systems having even degeneracy because odd electron
systems always belong to the previous case, by Kramers's theorem. We have
already proven that it will never be possible to have the spin operator satisfy
$X_S^*=\epsilon X_S$, no matter which basis is chosen. Nevertheless, as shown
by Griffith,\cite{Griffith1963} a simplification of the spin operator similar
to \eqnref{Qrules1} obtains if the basis consists of time conjugate pairs.
Given a real basis $\psi_i$ ($\psi_i^*=\psi_i$) (and this is always possible
for even-electron systems), construct a new basis $\psi(M)$ as follows:
\begin{equation*}
\begin{split}
\psi(\pm S)&=\frac{1}{\sqrt{2}}(\psi_1\pm i\psi_2),\\
\psi(\pm (S-1))&=\frac{1}{\sqrt{2}}(\psi_3\pm i\psi_4),
\end{split}
\end{equation*}
and so on. Then evidently $\psi(M)^*=\psi(-M)$, and it is easily verified that this
corresponds to $A_S=(-1)^{-S_z-S}Y_S$. From \eqnref{spinopTR} follows the
time-reversal behavior of $X_S$:
\begin{equation}\label{spinopTR2}
X_S^* = \epsilon(-1)^{S_z}X_S(-1)^{-S_z}
\end{equation}
Formally, $X_S$ undergoes a rotation through $\pi$ about the $z$ axis followed
by multiplication with $\epsilon$. To translate \eqnref{spinopTR2} into a rule
on the coefficients we use that $(-1)^{S_z}\ito{k}{q}(-1)^{-S_z}=
(-1)^q \ito{k}{q}$ and find\cite{Griffith1963} 
\begin{multline*}\label{Qrules2}
X_S^\dagger = X_S \text{ and } X_S^*=\epsilon(-1)^{S_z}
X_S(-1)^{-S_z} \quad \Leftrightarrow \\
Q_q^{(k)}=0 \text{ for}
\begin{cases}
 \text{$\epsilon=+1$ and $k-q$ odd}\\
 \text{$\epsilon=-1$ and $k-q$ even}
\end{cases}
\end{multline*}

\end{document}